\documentclass[prl,showpacs,amsmath,amssymb,twocolumn]{revtex4}

\usepackage{amsthm}
\usepackage{dcolumn}
\usepackage{bm}
\usepackage{graphicx}

\newtheorem{lemma}{Lemma}
\newtheorem{theorem}{Theorem}

\newcommand\ket[1]{\ensuremath{|#1\rangle}}
\newcommand\bra[1]{\ensuremath{\langle#1|}}
\newcommand\iprod[2]{\ensuremath{\langle#1|#2\rangle}}
\newcommand\oprod[2]{\ensuremath{|#1\rangle\langle#2|}}
\newcommand\tr{\mathop{\rm tr}\nolimits}

\begin{document}

\title{Identification and Distance Measures of Measurement Apparatus}

\author{Zhengfeng Ji}
  \email{jizhengfeng98@mails.tsinghua.edu.cn}
\author{Yuan Feng}
  \email{feng-y@tsinghua.edu.cn}
\author{Runyao Duan}
  \email{dry02@mails.tsinghua.edu.cn}
\author{Mingsheng Ying}
  \email{yingmsh@tsinghua.edu.cn}
\affiliation{
State Key Laboratory of Intelligent Technology and Systems, Department of Computer Science and Technology, Tsinghua University, Beijing 100084, China
}

\date{\today}

\begin{abstract}
We propose simple schemes that can perfectly identify projective measurement apparatus secretly chosen from a finite set. Entanglements are used in these schemes both to make possible the perfect identification and to improve the efficiency significantly. A brief discussion on the problem of how to appropriately define distance measures of measurements is also provided based on the results of identification.
\end{abstract}


\maketitle

Identification of physical objects, including both quantum states and quantum operations, has been an important subject of quantum information theory. State identification is extensively studied because it is found to be closely related to the fundamental feature of nonorthogonality of quantum mechanics. Especially, perfect identification is impossible for nonorthogonal states unless the number of copies of the unknown states goes infinite~\cite{DG98}. To one's surprise, things get quite different when we identify operations: it is always possible to completely tell apart any two different unitary operations by only finite number of uses of the unknown devices~\cite{Aci01,DPP01}. In this letter, we will show that projective measurements~\cite{Neu32}, the other most fundamental element of quantum mechanics, can also be distinguished with certainty despite the uncertain nature of quantum measurements.

Most of the previous works~\cite{Aci01,Che05,DSK05,DMS05,Yan05,WY06} reduce the problem of operation identification to that of state identification. We will take a more direct strategy in measurement identification by fully exploiting the outcome of the unknown device. Namely, instead of designing a new measurement as in both state identification and unitary identification, we make a clever use of the apparatus to be identified itself. The idea can be best illustrated by the following simple example. Suppose the observable to be identified is either $\sigma_z$ or $\sigma_x$, the Pauli matrices. The identification can be done by preparing $(\ket{00}-\ket{11})/\sqrt{2}$ and measuring both qubits with the unknown apparatus. It is easy to see that if the two results coincide, the apparatus is $\sigma_z$, otherwise $\sigma_x$. In the example, the measurement apparatus have proved their identities ``on their own'' without the help of any extra measurements and we will see that all measurements can be identified in this fashion.

Our next example employs the $n$-qubit $W$ state to witness the identity of an unknown observable which is known to be either $S=\sigma_z$ or $T=\oprod{\psi_0}{\psi_0}-\oprod{\psi_1}{\psi_1}$ where $\ket{\psi_0}=a\ket{0} + b\ket{1}$, $\ket{\psi_1}=b\ket{0} - a\ket{1}$ and $a=\sqrt{(n-1)/n}$, $b=1/\sqrt{n}$. As in the previous example, we measure all the $n$ qubits with the unknown observable. Simple calculation shows that only one of the $n$ outcomes is $-1$ if and only if the unknown device is $S$. Thus identification is done by simply counting the number of $-1$'s in the outcomes. In this example, the unknown apparatus are carried out exactly $n$ times and we will see later that a more efficient method exists. In fact, we will utilize the famous multiparticle entanglement, Greenberger-Horne-Zeilinger (GHZ) state~\cite{GHZ89}, to achieve optimal identification of single-qubit observables. Entanglement is thus as beneficial in improving distinguishability of measurements as it has been in varieties of other known applications~\cite{BW92,SJFT98,KF00,DPP01,ACS01}.

\begin{figure}[ht]
  \centering
  \includegraphics{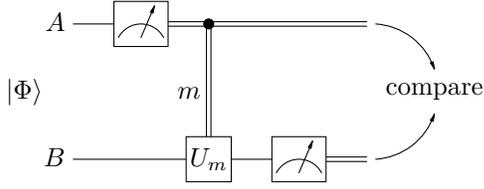}
  \caption{Illustration of the M--U--M scheme.}
  \label{fig:mum}
\end{figure}

More generally, the apparatus to be identified would be either $M = \sum_m m P_m$ or $N = \sum_m m Q_m$ where $P_m$, $Q_m$ are projectors of state space $\mathcal{H}$. We will focus on the problem of distinguishing two measurements and the general case can be dealt with in a similar way. Without loss of generality, $M$ and $N$ are assumed to have the same set of possible outcomes. Identification of this general problem needs the help of unitary operations. We demonstrate the idea in the example below and will extend it later to prove the general result. In this example, $M$ and $N$ of a $3$-level system are specified by $P_m=\oprod{m}{m}$ for $m=1,2,3$ and $Q_1$, $Q_2$, $Q_3$, three rank one projectors corresponding to $\ket{\psi_1} = (\ket{1}-2\ket{2}+\ket{3})/\sqrt{6}$, $\ket{\psi_2} = (\ket{1} + \ket{2} + \ket{3})/\sqrt{3}$, $\ket{\psi_3} = (\ket{1}-\ket{3})/\sqrt{2}$. We prepare a maximal entangled state $(\ket{11}+\ket{22}+\ket{33})/\sqrt{3}$ and measure the first qutrit labeled by $A$ as in Fig.~\ref{fig:mum}. Let the outcome be $1$. The state of the second qutrit is now either $\ket{1}$ or $\ket{\psi_1}$ depending on the unknown apparatus. Apply to it a unitary operation which keeps $\ket{1}$ unchanged and rotates $\ket{\psi_1}$ to a state orthogonal to itself. Such a unitary can be
\begin{equation*}
  U_1 = \frac{1}{5}
  \begin{bmatrix}
    5 & 0 & 0\\
    0 & -1 & \sqrt{24}\\
    0 & -\sqrt{24} & -1
  \end{bmatrix},
\end{equation*}
as $U_1\ket{1}=\ket{1}$ and $\bra{\psi_1} U_1 \ket{\psi_1} = 0$. If the second measurement still outputs $1$, the unknown device is definitely $M$, otherwise it is $N$. The case when the first outcome is other than $1$ can be solved similarly by choosing proper $U_2$ or $U_3$. We call such an identification strategy summarized in Fig.~\ref{fig:mum} M--U--M scheme.

{\em Identifying single-qubit observables.\/}---As single-qubit observables are the simplest and at the same time the most important, we will first discuss how to efficiently identify them and will come back to the general case later. In the introduction, we have shown how to do identification with $W$ state. The method used there is quite simple: we just prepare a pure state in $\mathcal{H}^{\otimes n}$ and measure $n$ times without performing any extra quantum operations. The decision is made depending solely on the $n$ measurement results. We call such an approach the simple scheme. We will also need another useful scheme, the M--M scheme. It is thus named because it simply modifies the simple scheme by allowing an extra known measurement to perform state identification after measuring the unknown apparatus. Namely, in this scheme, we prepare some pure state $\ket{\xi'}$ and measure the unknown measurement $n$ times. If the decision cannot be made yet, we discriminate the part of the state not measured. Only von Neumann measurements are considered in these two schemes; therefore we can write $P_m=\oprod{\phi_m}{\phi_m}$ and $Q_m=\oprod{\psi_m}{\psi_m}$. Denote by $U_M=\sum_i\oprod{\phi_i}{i}$, $U_N=\sum_i\oprod{\psi_i}{i}$ the associated unitary of $M$ and $N$ respectively and define the correlation unitary of $M$ and $N$ by $U = U_M^{\dagger}U_N=(\iprod{\phi_i}{\psi_j})$. We have the following theorems for these two schemes.

\begin{theorem}\label{thm:simple}
Let $M$ and $N$ be two von Neumann measurements and $U$ be their correlation unitary. $M$ and $N$ can be identified in the simple scheme within $n$ uses if and only if there exists some state $\ket{\xi} \in \mathcal{H}^{\otimes n}$ that nullifies the diagonal of $\oprod{\xi}{\xi}U^{\otimes n}$. The state used in the scheme can be $U_M^{\otimes n}\ket{\xi}$.
\end{theorem}

We omit the proof which follows easily from the proof of Theorem~\ref{thm:mm}. The criteria can in fact be further simplified.

\begin{theorem}\label{thm:submatrix}
$M$ and $N$ can be identified in the simple scheme within $n$ uses if and only if $U^{\otimes n}$ has a singular submatrix with the same row and column index set.
\end{theorem}

\begin{proof}
We prove the ``only if'' part first. Choose $\ket{\xi}$ such that $\oprod{\xi}{\xi} U^{\otimes n}$ has zero diagonal as Theorem~\ref{thm:simple} promised. Expand $\ket{\xi}$ in the computational basis as $\ket{\xi} = \sum_i a_i \ket{i}$. Denote $I=\{\, i \!\mid\! a_i \ne 0\,\}$. For all $i\in I$, we have $\iprod{i}{\xi}\bra{\xi} U^{\otimes n} \ket{i} = a_i \bra{\xi} U^{\otimes n} \ket{i} = 0$. Hence $\bra{\xi} U^{\otimes n} \ket{i} = 0$ which means that submatrix with rows and columns in set $I$ has an eigenvalue $0$ and is thus singular. The proof of the ``if'' part simply reverses the above procedure.
\end{proof}

\begin{theorem}\label{thm:mm}
$M$ and $N$ can be identified in the M--M scheme within $n$ uses if and only if there exists some density matrix $\rho$ that nullifies the diagonal of $\rho U^{\otimes n}$. The state used can be any purification of $U_M^{\otimes n}\rho U_M^{\dagger \otimes n}$.
\end{theorem}

\begin{proof}
We prove necessity only. Let $\ket{\xi'} \in \mathcal{H}_A \otimes \mathcal{H}_B$. $\mathcal{H}_A = \mathcal{H}^{\otimes n}$ is where the unknown measurement is performed. Expand $\ket{\xi'}$ as $\ket{\xi'} = \sum_i \ket{\phi_i} \ket{\hat{\xi}_i}$ and $\ket{\xi'} = \sum_i \ket{\psi_i} \ket{\tilde{\xi}_i}$ where $i=(i_1,i_2,\ldots,i_n)$ and $\ket{\phi_i}$ ($\ket{\psi_i}$) is the tensor product of $\ket{\phi_{i_j}}$ ($\ket{\psi_{i_j}}$). When observing result $i$ , perfect identification is possible only if (1) one of $\ket{\hat{\xi}_i}$ and $\ket{\tilde{\xi}_i}$ has zero norm and hence only one possible apparatus can have $i$ as its outcome or (2) $\ket{\hat{\xi}_i}$ and $\ket{\tilde{\xi}_i}$ are orthogonal and the following state identification can complete the decision. In both cases, we have $\iprod{\tilde{\xi}_i}{\hat{\xi}_i} = 0$. That is, for all $i$, $\bra{\xi'} U_N^{\otimes n} \oprod{i}{i} U_M^{\dagger \otimes n} \ket{\xi'} = 0$ or equivalently $\tr(\bra{i} U_M^{\dagger \otimes n} \oprod{\xi'}{\xi'} U_N^{\otimes n} \ket{i}) = 0$. Let $\rho' = \tr_B (\oprod{\xi'}{\xi'})$ and we get $\tr_A(\bra{i} U_M^{\dagger \otimes n} \rho' U_N^{\otimes n} \ket{i}) = 0$. Thus, $U_M^{\dagger \otimes n} \rho' U_N^{\otimes n}$ has zero diagonal. Finally, we can choose $\rho = U_M^{\dagger \otimes n} \rho'  U_M^{\otimes n}$ to complete the proof.
\end{proof}

To apply there results to the case of qubit observables, let the unknown observable be either $S = \oprod{\phi_0}{\phi_0} - \oprod{\phi_1}{\phi_1}$ or $T = \oprod{\psi_0}{\psi_0} - \oprod{\psi_1}{\psi_1}$. Noticing that the distinguishability of $S$ and $T$ is equal to that of $RSR^{\dagger}$ and $RTR^{\dagger}$ for any qubit rotation $R$, we can assume $S = \sigma_z$. Let $\ket{\psi_0} = \cos(\theta/2) \ket{0} + \sin(\theta/2) e^{i\varphi} \ket{1}$, $\ket{\psi_1} = \sin(\theta/2) \ket{0} - \cos(\theta/2) e^{i\varphi} \ket{1}$. Here, $\theta$ has a nice geometric interpretation---the angle between $\ket{\psi_0}$ and $\ket{0}$ in the Bloch sphere visualization~\cite{NC00}. For simplicity, sometimes we abbreviate $\cos(\theta/2)$ and $\sin(\theta/2)$ as $a$ and $b$ respectively. The correlation unitary $U$ of $S$ and $T$ is then $\begin{bmatrix} a & b\\ b e^{i\varphi} & -a e^{i\varphi} \end{bmatrix}$. As $\oprod{\xi}{\xi} U^{\otimes n}$ and $\oprod{\xi'}{\xi'} (VU)^{\otimes n}$ have zero diagonals simultaneously for $V=\oprod{0}{0} + e^{-i\varphi}\oprod{1}{1}$ and $\ket{\xi'} = V^{\otimes n} \ket{\xi}$, we can assume $\varphi=0$.  Denote by $w$ and $d$ the Hamming weight and Hamming distance function respectively. It is easily seen that $U^{\otimes n}$ is a $2^n \times 2^n$ matrix with the $(i,j)$-th element
\begin{equation}
  \label{eq:ijth}
  (-1)^{w(i\cdot j)} a^{n-d(i,j)} b^{d(i,j)},
\end{equation}
where $i,j$ are $n$-bit digits and $i\cdot j$ is their bit-wise and.

Our previous $W$ state example is implied by the fact that the submatrix of $U^{\otimes n}$ with index set $W_n=\{\, i \!\mid\! w(i) = 1 \,\}$ is singular when $\sin^2\frac{\theta}{2} = 1/n$. A more important example uses Theorem~\ref{thm:simple} with $\ket{\xi} = \sum_{i\in E_n} (-1)^{w(i)/2} \ket{i}/\sqrt{2^{n-1}}$ where the index set $E_n = \{\, i \!\mid\! w(i)\text{ is even}, 0 \le i < 2^n \,\}$. The $i$-th diagonal element of $\oprod{\xi}{\xi} U^{\otimes n}$ is obviously $0$ for $i\not\in E_n$. While for $i\in E_n$ it is
\begin{eqnarray}\label{eq:diag}
 &   & \frac{1}{2^{n-1}} \sum_{j\in E_n} (-1)^{w(i\cdot j)} a^{n-d(i,j)} b^{d(i,j)} (-1)^{(w(i)+w(j))/2}\nonumber\\
 & = & \frac{1}{2^{n-1}} \sum_{j\in E_n} (-1)^{d(i,j)/2} a^{n-d(i,j)} b^{d(i,j)}\nonumber\\
 & = & \frac{1}{2^{n-1}} \sum_{l=0}^n \sum_{j\in E_n, d(i,j)=l} (-1)^{l/2} a^{n-l} b^l\nonumber\\
 & = & \frac{1}{2^{n-1}} \sum_{l \text{ is even}} (-1)^{l/2} \binom{n}{l} \cos^{n-l}\frac{\theta}{2} \sin^l\frac{\theta}{2}\nonumber\\
 & = & \frac{1}{2^{n-1}} \cos\frac{n\theta}{2}.
\end{eqnarray}
The first equality follows from the fact that $w(i) + w(j) = d(i,j) + 2w(i\cdot j)$ since both sides count the number of 1's in $i, j$. The last one follows from the de Moivre's identity. When $\theta = \frac{2k+1}{n}\pi$, all elements on the diagonal of $\oprod{\xi}{\xi} U^{\otimes n}$ become zero and it follows from Theorem~\ref{thm:simple} that $S$ and $T$ can be identified by the simple scheme with $n$ uses of the apparatus. The entanglement we use in this scheme is $\ket{G_n} = \sum_{i\in E_n} (-1)^{w(i)/2}\ket{i} / \sqrt{2^{n-1}}$. Such a state can be efficiently generated by performing parity measurement~\cite{ZZY05} on state $\frac{\ket{0}+i\ket{1}}{\sqrt{2}}^{\otimes n}$ and it is not difficult to see that $\ket{G_n}$ is equivalent to the GHZ state up to local unitaries.

For many different $\theta$, $S$ and $T$ can be identified in the simple scheme as we have already shown. However, for all such $\theta$, $\tan\theta$ is an algebraic number satisfying an integer-coefficient polynomial equation which follows from the zero-determinant property of the corresponding singular submatrix. This means that simple identification is generally impossible and causes the most severe drawback of the simple scheme. Fortunately, the M--M scheme solves this problem and we can in fact prove that for any $\theta$ the optimal M--M scheme measures the unknown apparatus $\lceil \pi/\theta \rceil$ times.

The construction goes as follows. Let $n=\lceil \pi/\theta \rceil$. Our aim is to find some $\rho$ such that $\rho U^{\otimes n}$ has zero diagonal as Theorem~\ref{thm:mm} guarantees. If we simply set $\rho=\oprod{G_n}{G_n}$, Eq.~\eqref{eq:diag} indicates that half of the diagonal elements are already $0$ while the other half, with index in $E_n$, are the same number $\frac{1}{2^{n-1}} \cos\frac{n\theta}{2}$. They are negative as $n=\lceil \pi/\theta \rceil$. We need to fix the negative half. To this end, partition $E_n$ into $E_n^1,E_n^2,\ldots,E_n^n$ where $E_n^1$ contains all binary integers in $E_n$ that end with $0$, $E_n^2$ contains those having suffix $11$, $E_n^3$ having suffix $101$ and so on. Next, construct a series of states as $\ket{G_n^1} = \ket{G_{n-1}}\ket{0}$, $\ket{G_n^2} = \ket{G_{n-2}}\ket{11}$, $\ket{G_n^3} = \ket{G_{n-3}}\ket{101}$,$\ldots$,$\ket{G_n^{n-1}} = \ket{G_1}\ket{1}\ket{0}^{n-3}\ket{1}$,$\ket{G_n^n} = \ket{1}\ket{0}^{n-2}\ket{1}$. Simple calculation gives that $\oprod{G_n^i}{G_n^i} U^{\otimes n}$ has nonzero diagonal elements only when the index is in $E_n^i$ and these nonzero elements have a same positive value. Hence, by properly choosing a probability distribution over $\ket{G_n}$ and $\ket{G_n^i}$ we can have an appealing ensemble required in Theorem~\ref{thm:mm}.

The proof of optimality is somewhat easier. Suppose we can identify $S$ and $T$ by the M--M scheme within $n$ uses. Theorem~\ref{thm:mm} guarantees that there exists some $\rho$ such that $\rho U^{\otimes n}$ has zero diagonal. Obviously, $\rho U^{\otimes n} \sigma_z^{\otimes n}$ also has zero diagonal and thus has zero trace. Results from Ref.~\cite{Aci01} insure $n \ge \lceil \pi/\theta \rceil$, as $U \sigma_z$ has eigenvalues $e^{\pm i \theta/2}$.

One thing worth noting is that the extra known measurement in the M--M scheme can be replace by a unitary operation. Suppose $S$ and $T$ can be identified using the M--M scheme by measuring the unknown apparatus $n$ times. Let $\ket{\xi_S}$ ($\ket{\xi_T}$) be the state left after measuring $S$ ($T$) $n-1$ times. Write $\ket{\xi_S} = \ket{0} \ket{\alpha_0} + \ket{1} \ket{\alpha_1}$ and $\ket{\xi_T} = \ket{\psi_0} \ket{\beta_0} + \ket{\psi_1} \ket{\beta_1}$. The property of M--M scheme indicates $\iprod{\alpha_i}{\beta_i} = 0$ for $i=1,2$. Using Cauchy-Schwarz inequality twice, we have $|\iprod{\xi_S}{\xi_T}| \le b (|\iprod{\alpha_0}{\beta_1}| + |\iprod{\alpha_1}{\beta_0}|) \le b (\left\|\ket{\alpha_0}\right\| \left\|\ket{\beta_1}\right\| + \left\|\ket{\alpha_1}\right\| \left\|\ket{\beta_0}\right\|) \le b$. A unitary operation $V$ can thus be chosen such that $V\ket{\xi_S} = \ket{0}\ket{\xi'_S}$ and $V\ket{\xi_T} = \ket{\psi_1}\ket{\xi'_T}$ for some $\ket{\xi'_S}$ and $\ket{\xi'_T}$. After applying $V$, we can measure the unknown apparatus for the last time and no state identification is necessary anymore. Using similar techniques, a general lower bound $O(1/\theta)$ can be proved for all possible identification schemes. Thus our M--M scheme is also asymptotically optimal in the most general setting.

{\em Identifying projective measurements.\/}---We now deal with the general case where $M=\sum_m mP_m$, $N=\sum_m mQ_m$. Use $P$ to represent also the corresponding projective subspace of a projector $P$ since they are one-to-one. The following lemma is needed to construct the general identification scheme.

\begin{lemma}\label{lem:sep}
Let $P$ and $Q$ be two projectors on $d$ dimensional space $\mathcal{H}$. Ranks of $P$ and $Q$ are both $r$. Then there exists some unitary $U$ such that $UP^*U^{\dagger}=P$ and $UQ^*U^{\dagger} \perp Q$ if $\|PQ\| \le 1/\sqrt{2}$ and $d \ge 3r$ where $P^*$ ($Q^*$) is the complex conjugate of $P$ ($Q$) and $\|\cdot\|$ is the operator norm. We call $U$ the separation unitary of $P$ and $Q$.
\end{lemma}

\begin{proof}
Expand $P = \sum_{i=1}^r\limits \oprod{\phi_i}{\phi_i}$ and $Q = \sum_{i=1}^r\limits \oprod{\psi_i}{\psi_i}$. As $\|PQ\| < 1$, we have $P \cap Q = \{0\}$ and therefore the dimension of $span(P, Q)$ is $2r$. Let $\ket{\psi_j} = \sum_{i=1}^r a_{i,j}\ket{\phi_i} + \sum_{i=1}^r b_{i,j}\ket{\xi_i}$ where $\{\ket{\xi_i}\}$ together with $\{\ket{\phi_i}\}$ form an orthonormal basis of $span(P, Q)$. Let $A = (a_{i,j})$ and $B = (b_{i,j})$. It follows from the orthonormal property of $\{\ket{\psi_j}\}$ that $A^{\dagger}A+B^{\dagger}B=I$ and from the unitarily invariant property of operator norm that $\|A\| = \| PQ\|$.

First, we choose $U$ such that $U\ket{\phi_i^*} = \ket{\phi_i}$ for all $i=1,\ldots,r$, then $UP^*U^{\dagger}=P$ is obviously satisfied. If we extend $\{\ket{\phi_i},\ket{\xi_i}\}$ to a complete basis $\{\ket{\omega_i}\}$ of $\mathcal{H}$ and write out the matrix representation of $U$ with respect to $\{\ket{\omega_i^*}\}$ and $\{\ket{\omega_i}\}$ of the input and output spaces respectively, then $U$ is in fact chosen to have a blocked form like
\begin{equation}\label{eq:block}
  U =
  \begin{bmatrix}
    I & & \\
      & V & \cdots\\
      & \vdots & \ddots
  \end{bmatrix}
\end{equation}
where $V$ is an $r$ by $r$ matrix with $V_{i,j}=\bra{\xi_i}U\ket{\xi_j^*}$. The second requirement, $UQ^*U^{\dagger} \perp Q$, is equivalent to $\bra{\psi_i}U\ket{\psi_j^*} = 0$ for all $i,j$ and it can be further simplified to $A^{\dagger}A^* + B^{\dagger}VB^* = 0$. Thus $V=-(B^{\dagger})^{-1}A^{\dagger}A^*(B^*)^{-1}$ and
\begin{equation*}
\begin{split}
  \|V\| & \le \|A\|^2\,\|B^{-1}\|^2\\
        & =   \|A\|^2\,\|(B^{\dagger}B)^{-1}\|\\
        & =   \|A\|^2\,\|(I-A^{\dagger}A)^{-1}\|\\
        & \le \|A\|^2\sum_{i=0}^{\infty} \|A^{\dagger}A\|^i\\
        & =   \|A\|^2\frac{1}{1-\|A\|^2} = \frac{\|PQ\|^2}{1-\|PQ\|^2} \le 1.
\end{split}
\end{equation*}
The second inequality is the triangle inequality applied to the Neumann series $(I-N)^{-1} = I + N + N^2 + \cdots$ when $\|N\| < 1$. Employing an exercise in Ref.~\cite{Bha96}, we know that $U$ of form Eq.~\eqref{eq:block} can be extended to a unitary as $\|V\| \le 1$ and $d \ge 3r$.
\end{proof}

If $P_m$ and $Q_m$ satisfy the conditions in the above lemma for all $m$, we can just follow the M--U--M scheme depicted in Fig.~\ref{fig:mum} where $U_m$ is the separation unitary of $P_m$ and $Q_m$ as Lemma~\ref{lem:sep} guarantees. To see that $M$ always leads to the same results, we use the identity
\begin{equation*}
  P_m \otimes I \ket{\Phi} = I \otimes P_m^* \ket{\Phi}.
\end{equation*}
Therefore, after $U_m$ is applied, the state becomes $I \otimes U_mP_m^* \ket{\Phi}$ without normalization and is equal to $I \otimes P_mU_m \ket{\Phi}$. Due to the repeatability of projective measurements, the second measurement will always get $m$. The fact that observable $N$ leads to two different outcomes follows from a similar argument.

We now deal with the case when conditions in the lemma are not satisfied. The first possibility is that some $P_m$ and $Q_m$ have different ranks, for example, $rank(P_m) > rank(Q_m)$. Then we are able to find some $\ket{\phi} \in P_m$ such that $\ket{\phi} \perp Q_m$. The unknown apparatus can be identified by simply preparing state $\ket{\phi}$ and measuring it. Secondly, if for all $m$, $P_m$ and $Q_m$ have the same rank and $\|P_mQ_m\| < 1$, consider multiple measurement in parallel, namely $M^{\otimes L}$ or $N^{\otimes L}$. For sufficient large $L$, both the norm condition $\|PQ\|\le 1/\sqrt{2}$ and the dimensionality condition $d \ge 3r$ can be satisfied. The last special case left is $\|P_mQ_m\| = 1$ for some $P_m$ and $Q_m$. We reduce it to the previous case by noticing that such an unknown apparatus can simulate another unknown measurement whose projective subspaces are $P'_m$ or $Q'_m$ where $P'_m = P_m \cap (P_m \cap Q_m)^{\perp}$, $Q'_m = Q_m \cap (P_m \cap Q_m)^{\perp}$ and $\|P'_mQ'_m\|<1$.

It is worth noting that, in the M--U--M scheme, no post-measurement states are used for further processing. However, if an experiment permits further manipulation of post-measurement states, we can even replace the bipartite entanglement with an arbitrary state and apply on it all the operations, the two measurements and the unitary operation in between.

{\em Distance measures of measurements.\/}---When we want to quantify how different two observables are, the first idea that comes into mind might be to compare their probabilistic behavior. Namely, we can define
\begin{equation}
  \label{eq:dmax}
  D_{max}(M, N) = \sup_{\rho} D(p_m, q_m),
\end{equation}
where $p_m = \tr(\rho P_m)$ and $q_m = \tr(\rho Q_m)$. This definition is in some sense the dual of the trace distance for density operators as Theorem 9.1 of Ref.~\cite {NC00} indicates. However, it is not a good definition in general except for single-qubit observables. For one thing, Theorem~\ref{thm:submatrix} guarantees that $D_{max}(S^{\otimes 2}, T^{\otimes 2}) = 1$ only if $\theta$ is $\pi$ or $\pi/2$. This can be shown by checking submatrices of $U^{\otimes 2}$. It contradicts our intuition that the larger the value of $\theta$ the more different the observables. What is more, this definition violates the stable requirement of measures for operations~\cite{Kit97,GLN05}, $\Delta(\mathcal{E},\mathcal{F})=\Delta(\mathcal{I}\otimes\mathcal{E}, \mathcal{I}\otimes\mathcal{F})$. Indeed, if we think of $S$ and $T$ as quantum operations with Kraus representation $\{ \oprod{0}{0}, \oprod{1}{1} \}$ and $\{ \oprod{0}{\psi_0}, \oprod{1}{\psi_1} \}$ respectively, then $D_{max}(S,T)$ defined above is exactly the same as it is in Ref.~\cite{GLN05} and is already known to be problematic. To solve this problem, we can use, for example, the stabilized version $D_{stab}=D_{max}(\mathcal{I}\otimes\mathcal{E}, \mathcal{I}\otimes\mathcal{F})$ as Ref.~\cite{GLN05} recommended.

It follows from Theorem~\ref{thm:mm} that $D_{stab}(S^{\otimes 2}, T^{\otimes 2})=1$ for $\theta \ge \pi/2$ and from Theorem~\ref{thm:submatrix} that $D_{max}(S^{\otimes 2}, T^{\otimes 2}) < 1$ for all $\theta \in (\pi/2, \pi)$. Thus $D_{stab}$ and $D_{max}$ are different for multi-qubit measurements. Yet, for single qubit observables $S$ and $T$, $D_{stab}$ can be calculated explicitly and turns out to be equal to $D_{max} = \sin\frac{\theta}{2}$. Fidelity of observables can be similarly studied and $F_{stab}$ is generally not equal to $F_{min}$ except for the qubit case where $F_{stab} = F_{min} = \cos\frac{\theta}{2}$. It is somewhat strange that only single-qubit observables can be differentiated using the intuitive approach.

To summarize, we have proved that all projective measurements are distinguishable and have found the optimal method for identifying qubit observables. We conclude that both nonorthogonality and uncertainty in projective measurements can be tackled to achieve perfect identification. As an application, definitions of distance measures of measurements are briefly discussed and it is found that probabilistic behavior is generally incapable of fully differentiating quantum measurements.

We are thankful to the colleagues in the Quantum Computation and Information Research Group for helpful discussions. This work was partly supported by the National Natural Science Foundation of China (Grant Nos. 60503001, 60321002, and 60305005), and by Tsinghua Basic Research Foundation (Grant No. 052220204). R. Duan acknowledges the financial support of Tsinghua University (Grant No. 052420003).

\bibliography{imoma}

\begin{thebibliography}{19}
\expandafter\ifx\csname natexlab\endcsname\relax\def\natexlab#1{#1}\fi
\expandafter\ifx\csname bibnamefont\endcsname\relax
  \def\bibnamefont#1{#1}\fi
\expandafter\ifx\csname bibfnamefont\endcsname\relax
  \def\bibfnamefont#1{#1}\fi
\expandafter\ifx\csname citenamefont\endcsname\relax
  \def\citenamefont#1{#1}\fi
\expandafter\ifx\csname url\endcsname\relax
  \def\url#1{\texttt{#1}}\fi
\expandafter\ifx\csname urlprefix\endcsname\relax\def\urlprefix{URL }\fi
\providecommand{\bibinfo}[2]{#2}
\providecommand{\eprint}[2][]{\url{#2}}

\bibitem[{\citenamefont{Duan and Guo}(1998)}]{DG98}
\bibinfo{author}{\bibfnamefont{L.-M.} \bibnamefont{Duan}} \bibnamefont{and}
  \bibinfo{author}{\bibfnamefont{G.-C.} \bibnamefont{Guo}},
  \bibinfo{journal}{Physical Review Letters} \textbf{\bibinfo{volume}{80}},
  \bibinfo{pages}{4999} (\bibinfo{year}{1998}),
  \urlprefix\url{http://link.aps.org/abstract/PRL/v80/p4999}.

\bibitem[{\citenamefont{Acin}(2001)}]{Aci01}
\bibinfo{author}{\bibfnamefont{A.}~\bibnamefont{Acin}},
  \bibinfo{journal}{Physical Review Letters} \textbf{\bibinfo{volume}{87}},
  \bibinfo{eid}{177901} (pages~\bibinfo{numpages}{4}) (\bibinfo{year}{2001}),
  \urlprefix\url{http://link.aps.org/abstract/PRL/v87/e177901}.

\bibitem[{\citenamefont{D'Ariano et~al.}(2001)\citenamefont{D'Ariano, Presti,
  and Paris}}]{DPP01}
\bibinfo{author}{\bibfnamefont{G.~M.} \bibnamefont{D'Ariano}},
  \bibinfo{author}{\bibfnamefont{P.~L.} \bibnamefont{Presti}},
  \bibnamefont{and} \bibinfo{author}{\bibfnamefont{M.~G.~A.}
  \bibnamefont{Paris}}, \bibinfo{journal}{Physical Review Letters}
  \textbf{\bibinfo{volume}{87}}, \bibinfo{eid}{270404}
  (pages~\bibinfo{numpages}{4}) (\bibinfo{year}{2001}),
  \urlprefix\url{http://link.aps.org/abstract/PRL/v87/e270404}.

\bibitem[{\citenamefont{von Neumann}(1932)}]{Neu32}
\bibinfo{author}{\bibfnamefont{J.}~\bibnamefont{von Neumann}},
  \emph{\bibinfo{title}{Mathematische Grundlagen der Quantenmechanik}}
  (\bibinfo{publisher}{Springer, Berlin}, \bibinfo{year}{1932}).

\bibitem[{\citenamefont{Chefles}(2005)}]{Che05}
\bibinfo{author}{\bibfnamefont{A.}~\bibnamefont{Chefles}},
  \bibinfo{journal}{Physical Review A (Atomic, Molecular, and Optical Physics)}
  \textbf{\bibinfo{volume}{72}}, \bibinfo{eid}{042332}
  (pages~\bibinfo{numpages}{10}) (\bibinfo{year}{2005}),
  \urlprefix\url{http://link.aps.org/abstract/PRA/v72/e042332}.

\bibitem[{\citenamefont{D'Ariano
  et~al.}(2005{\natexlab{a}})\citenamefont{D'Ariano, Sacchi, and Kahn}}]{DSK05}
\bibinfo{author}{\bibfnamefont{G.~M.} \bibnamefont{D'Ariano}},
  \bibinfo{author}{\bibfnamefont{M.~F.} \bibnamefont{Sacchi}},
  \bibnamefont{and} \bibinfo{author}{\bibfnamefont{J.}~\bibnamefont{Kahn}},
  \bibinfo{journal}{Physical Review A (Atomic, Molecular, and Optical Physics)}
  \textbf{\bibinfo{volume}{72}}, \bibinfo{eid}{052302}
  (pages~\bibinfo{numpages}{7}) (\bibinfo{year}{2005}{\natexlab{a}}),
  \urlprefix\url{http://link.aps.org/abstract/PRA/v72/e052302}.

\bibitem[{\citenamefont{D'Ariano
  et~al.}(2005{\natexlab{b}})\citenamefont{D'Ariano, Mataloni, and
  Sacchi}}]{DMS05}
\bibinfo{author}{\bibfnamefont{G.~M.} \bibnamefont{D'Ariano}},
  \bibinfo{author}{\bibfnamefont{P.}~\bibnamefont{Mataloni}}, \bibnamefont{and}
  \bibinfo{author}{\bibfnamefont{M.~F.} \bibnamefont{Sacchi}},
  \bibinfo{journal}{Physical Review A (Atomic, Molecular, and Optical Physics)}
  \textbf{\bibinfo{volume}{71}}, \bibinfo{eid}{062337}
  (pages~\bibinfo{numpages}{4}) (\bibinfo{year}{2005}{\natexlab{b}}),
  \urlprefix\url{http://link.aps.org/abstract/PRA/v71/e062337}.

\bibitem[{\citenamefont{Yang}(2005)}]{Yan05}
\bibinfo{author}{\bibfnamefont{D.}~\bibnamefont{Yang}},
  \emph{\bibinfo{title}{Distinguishability, classical information of quantum
  operations}} (\bibinfo{year}{2005}), \eprint{quant-ph/0504073},
  \urlprefix\url{http://arxiv.org/abs/quant-ph/0504073}.

\bibitem[{\citenamefont{Wang and Ying}(2006)}]{WY06}
\bibinfo{author}{\bibfnamefont{G.}~\bibnamefont{Wang}} \bibnamefont{and}
  \bibinfo{author}{\bibfnamefont{M.}~\bibnamefont{Ying}},
  \bibinfo{journal}{Physical Review A (Atomic, Molecular, and Optical Physics)}
  \textbf{\bibinfo{volume}{73}}, \bibinfo{eid}{042301}
  (pages~\bibinfo{numpages}{5}) (\bibinfo{year}{2006}),
  \urlprefix\url{http://link.aps.org/abstract/PRA/v73/e042301}.

\bibitem[{\citenamefont{Greenberger et~al.}(1989)\citenamefont{Greenberger,
  Horne, and Zeilinger}}]{GHZ89}
\bibinfo{author}{\bibfnamefont{D.~M.} \bibnamefont{Greenberger}},
  \bibinfo{author}{\bibfnamefont{M.~A.} \bibnamefont{Horne}}, \bibnamefont{and}
  \bibinfo{author}{\bibfnamefont{A.}~\bibnamefont{Zeilinger}},
  \emph{\bibinfo{title}{Bell's Theorem, Quantum Theory and Conceptions of the
  Universe}} (\bibinfo{publisher}{Kluwer Academic, Dordrecht},
  \bibinfo{year}{1989}).

\bibitem[{\citenamefont{Bennett and Wiesner}(1992)}]{BW92}
\bibinfo{author}{\bibfnamefont{C.~H.} \bibnamefont{Bennett}} \bibnamefont{and}
  \bibinfo{author}{\bibfnamefont{S.~J.} \bibnamefont{Wiesner}},
  \bibinfo{journal}{Physical Review Letters} \textbf{\bibinfo{volume}{69}},
  \bibinfo{pages}{2881} (\bibinfo{year}{1992}),
  \urlprefix\url{http://link.aps.org/abstract/PRL/v69/p2881}.

\bibitem[{\citenamefont{Saleh et~al.}(1998)\citenamefont{Saleh, Jost, Fei, and
  Teich}}]{SJFT98}
\bibinfo{author}{\bibfnamefont{B.~E.~A.} \bibnamefont{Saleh}},
  \bibinfo{author}{\bibfnamefont{B.~M.} \bibnamefont{Jost}},
  \bibinfo{author}{\bibfnamefont{H.-B.} \bibnamefont{Fei}}, \bibnamefont{and}
  \bibinfo{author}{\bibfnamefont{M.~C.} \bibnamefont{Teich}},
  \bibinfo{journal}{Physical Review Letters} \textbf{\bibinfo{volume}{80}},
  \bibinfo{pages}{3483} (\bibinfo{year}{1998}),
  \urlprefix\url{http://link.aps.org/abstract/PRL/v80/p3483}.

\bibitem[{\citenamefont{Kolobov and Fabre}(2000)}]{KF00}
\bibinfo{author}{\bibfnamefont{M.~I.} \bibnamefont{Kolobov}} \bibnamefont{and}
  \bibinfo{author}{\bibfnamefont{C.}~\bibnamefont{Fabre}},
  \bibinfo{journal}{Physical Review Letters} \textbf{\bibinfo{volume}{85}},
  \bibinfo{pages}{3789} (\bibinfo{year}{2000}),
  \urlprefix\url{http://link.aps.org/abstract/PRL/v85/p3789}.

\bibitem[{\citenamefont{D'Angelo et~al.}(2001)\citenamefont{D'Angelo, Chekhova,
  and Shih}}]{ACS01}
\bibinfo{author}{\bibfnamefont{M.}~\bibnamefont{D'Angelo}},
  \bibinfo{author}{\bibfnamefont{M.~V.} \bibnamefont{Chekhova}},
  \bibnamefont{and} \bibinfo{author}{\bibfnamefont{Y.}~\bibnamefont{Shih}},
  \bibinfo{journal}{Physical Review Letters} \textbf{\bibinfo{volume}{87}},
  \bibinfo{eid}{013602} (pages~\bibinfo{numpages}{4}) (\bibinfo{year}{2001}),
  \urlprefix\url{http://link.aps.org/abstract/PRL/v87/e013602}.

\bibitem[{\citenamefont{Nielsen and Chuang}(2000)}]{NC00}
\bibinfo{author}{\bibfnamefont{M.~A.} \bibnamefont{Nielsen}} \bibnamefont{and}
  \bibinfo{author}{\bibfnamefont{I.~L.} \bibnamefont{Chuang}},
  \emph{\bibinfo{title}{Quantum Computation and Quantum Information}}
  (\bibinfo{publisher}{Cambridge University Press}, \bibinfo{year}{2000}).

\bibitem[{\citenamefont{Zeng et~al.}(2005)\citenamefont{Zeng, Zhou, and
  You}}]{ZZY05}
\bibinfo{author}{\bibfnamefont{B.}~\bibnamefont{Zeng}},
  \bibinfo{author}{\bibfnamefont{D.~L.} \bibnamefont{Zhou}}, \bibnamefont{and}
  \bibinfo{author}{\bibfnamefont{L.}~\bibnamefont{You}},
  \bibinfo{journal}{Physical Review Letters} \textbf{\bibinfo{volume}{95}},
  \bibinfo{eid}{110502} (pages~\bibinfo{numpages}{4}) (\bibinfo{year}{2005}),
  \urlprefix\url{http://link.aps.org/abstract/PRL/v95/e110502}.

\bibitem[{\citenamefont{Bhatia}(1996)}]{Bha96}
\bibinfo{author}{\bibfnamefont{R.}~\bibnamefont{Bhatia}},
  \emph{\bibinfo{title}{Matrix Analysis}} (\bibinfo{publisher}{Springer},
  \bibinfo{year}{1996}), p.~\bibinfo{pages}{11}.

\bibitem[{\citenamefont{Kitaev}(1997)}]{Kit97}
\bibinfo{author}{\bibfnamefont{A.~Y.} \bibnamefont{Kitaev}},
  \bibinfo{journal}{Russian Mathematical Surveys}
  \textbf{\bibinfo{volume}{52}}, \bibinfo{pages}{1191} (\bibinfo{year}{1997}).

\bibitem[{\citenamefont{Gilchrist et~al.}(2005)\citenamefont{Gilchrist,
  Langford, and Nielsen}}]{GLN05}
\bibinfo{author}{\bibfnamefont{A.}~\bibnamefont{Gilchrist}},
  \bibinfo{author}{\bibfnamefont{N.~K.} \bibnamefont{Langford}},
  \bibnamefont{and} \bibinfo{author}{\bibfnamefont{M.~A.}
  \bibnamefont{Nielsen}}, \bibinfo{journal}{Physical Review A (Atomic,
  Molecular, and Optical Physics)} \textbf{\bibinfo{volume}{71}},
  \bibinfo{eid}{062310} (pages~\bibinfo{numpages}{14}) (\bibinfo{year}{2005}),
  \urlprefix\url{http://link.aps.org/abstract/PRA/v71/e062310}.

\end{thebibliography}

\end{document}